\documentclass[letterpaper]{JHEP3}
\usepackage[T1]{fontenc}
\usepackage[latin1]{inputenc}
\usepackage{amsmath}
\makeatletter
\usepackage{mathrsfs}
\author{P. Bantay%
\footnote{Institute for Theoretical Physics, E\"otv\"os University, Budapest}%
\quad and T. Gannon\footnote{Department of Mathematical Sciences, University of Alberta, Edmonton}}
\title{Conformal characters and the modular representation}
\abstract{A general procedure is presented to determine, given any
 suitable representation of the modular group, the characters of all possible Rational Conformal Field Theories whose associated modular representation is the given one. The relevant ideas and methods are illustrated on two non-trivial examples: the Yang-Lee and the Ising models.}

\makeatother

\begin{document}

%\contentsline {section}{\numberline {1}Introduction}{1}
%\contentsline {section}{\numberline {2}Admissible representations and the canonical basis}{2}
%\contentsline {section}{\numberline {3}The Hauptmodul and the recursion relations}{4}
%\contentsline {section}{\numberline {4}Eisenstein series and the differential relations}{5}
%\contentsline {section}{\numberline {5}Invariants and covariants}{8}
%\contentsline {section}{\numberline {6}A worked-out example: the Yang-Lee model}{9}
%\contentsline {section}{\numberline {7}A second example: the Ising model}{12}
%\contentsline {section}{\numberline {8}Further questions and developments}{15}

\newcommand{\mr}[4]{\rho\!\left(
\begin {array}{cc}
#1  &  #2\\
#3  &  #4 \end{array} \right)}
\newcommand{\set}[2]{\left\{  #1\, |\, #2\right\}  }
\newcommand{\aut}[1]{\mathrm{Aut}\!\left(#1\right)}
\newcommand{\sm}[4]{\left(\begin{smallmatrix}#1  &  #2\cr\cr#3  &  #4\end{smallmatrix}\right)}
\newcommand{\cyc}[1]{\mathbb{Q}\left[\zeta_{#1}\right]}
\newcommand{\Mod}[3]{#1\equiv#2\, \left(\mathrm{mod}\, \, #3\right)}
\newcommand{\SL}{\mathrm{SL}_{2}\!\left(\mathbb{Z}\right)}
\newcommand{\gl}[2]{\mathrm{GL}\!\left(#1,#2\right)}
\newcommand{\tr}[2]{\textrm{Tr}_{_{#1}}\left(#2\right)}
\newcommand{\FA}{\mathbb{X}}
\newcommand{\FB}{\Lambda}
\newcommand{\FC}{\mathbf{H}}
\newcommand{\FD}{\mathcal{M}\!\left(\rho\right)}
\newcommand{\FE}{\mathscr{P}}
\newcommand{\FF}{W_{\rho}}
\newcommand{\inv}{Inv\left(G\right)}
\newcommand{\ks}{\lambda}
\newcommand{\SB}[2]{\FA^{\left(#1;#2\right)}}
\newcommand{\w}[1]{\mathfrak{f}_{#1}}
\newcommand{\ev}[1]{{\bf e}_{#1}}
\newcommand{\pcsg}[1]{\Gamma\left(#1\right)}
\newcommand{\ccorb}{\mathcal{O}}
\newcommand{\maxid}{\mathfrak{m}}
\newcommand{\orb}{\eta}
\newcommand{\DO}{\nabla}
\newcommand{\ei}{\mathcal{E}}
\newcommand{\xm}[3]{\mathcal{X}_{#1}^{\left(#2;#3\right)}}

\section{Introduction}

Among the many quantities that characterize a Conformal Field Theory,
a special role is played by the characters $\chi_{\mu}\left(\tau\right)$
of the primary fields. By determining the degeneracies of the Virasoro
generator $L_{0}$ in the different sectors, they convey information
about how the chiral algebra of the theory is represented on the space
of states. As such, they are the basic building blocks of the torus
partition function, and from their knowledge one can read off at once
the conformal weights $h_{\mu}$ and the central charge $c$ of the
theory. Even more, they determine almost uniquely the modular $S$-matrix
of the theory, and thus the fusion rules via Verlinde's formula.

Reversing the logic, one may ask to what extent can the characters
be recovered from the knowledge of the conformal weights, central
charge and fusion rules. This paper aims to answer that question.
An elementary observation is that the fusion rules uniquely determine
$S$, up to at most a permutation of its columns and a multiplication of each
column by $\pm1$ \cite{De}. Together, $h_{\mu},\, c$ and $S$ uniquely determine
a representation $\rho$ of the modular group $\SL$. The characters
$\chi_{\mu}$ may be grouped into a {\it character vector} $\FA$ that is
holomorphic in the upper half-plane, transforms according to $\rho$,
and - in a suitable sense, to be explained in Section 2 - has only
finite order poles at $\tau=i\infty$. Thus the real question is:
to what extent does the modular representation $\rho$ determine the
characters? We explain that the characters of an RCFT are uniquely determined
by its modular representation and the singular terms $\sum_{s<0}a_sq^s$
of each character, up to perhaps some constant terms. We show  how
to construct
all character vectors compatible with a representation  $\rho$ that satisfies some simple conditions holding in any RCFT.

This paper collects our basic results, illustrating them with examples.
The follow-up paper \cite{BG} will give more details and push our analysis much
further.
Section 2 introduces the notation and establishes the extent to which
the characters are determined by $\rho$. Sections 3 to 5 describe
how to explicitly build the character vectors term-by-term, starting
from $\rho$. We conclude the paper with two concrete examples: the
Yang-Lee and Ising models. In the explicit computations, we made heavy
use of the Computer Algebra Systems GAP \cite{GAP} (for the computations
of invariants and covariants), PARI/GP \cite{PARI2} (for the computations
involving modular forms) and Singular \cite{GPS05} (for the solution
of polynomial systems).

\section{Admissible representations and the canonical basis}

Let's consider a finite dimensional (unitary) representation $\rho:\SL\rightarrow\gl{r}{\mathbb{C}}$
of the modular group $\SL$. It is known that such a representation
is completely characterized by the pair of matrices $T=\rho\!\sm{1}{1}{0}{1}$
and $S=\rho\!\sm{0}{-1}{1}{0}$, which satisfy the relations $S^{4}=1$
and $STS=T^{-1}ST^{-1}$. As usual, $\pcsg{N}$ will denote the {\it principal
congruence subgroup} of level $N$, i.e. \begin{equation}
\pcsg{N}=\set{\sm{a}{b}{c}{d}\in\SL}{\Mod{a,d}{1}{N}\,\,\mathrm{and}\,\,\Mod{b,c}{0}{N}}\,\,.\label{eq:pcsg}\end{equation}
 We'll denote by $\FF$ the subspace of $\mathbb{C}^{r}$ consisting
of those vectors that are invariant under $\rho$, i.e. $\FF=\set{v\in\mathbb{C}^{r}}{Tv=Sv=v}$,
by $\left\langle .,.\right\rangle $ the standard scalar product on
$\mathbb{C}^{r}$, and by $\ev{\mu}$ the $\mu$-th element of the
standard basis of $\mathbb{C}^{r}$, whose $\mu$-th entry is 1, and
all others 0.

We'll call a representation $\rho:\SL\rightarrow\gl{r}{\mathbb{C}}$
{\it admissible} if it satisfies the following conditions:

\begin{description}
\item [1]$\ker\rho$ is a {\it congruence subgroup}, i.e. it %$\Gamma\left(N\right)<\ker\rho$
contains $\Gamma\left(N\right)$ for some integer $N$;
\item [2]$T$ is diagonal and $S^{2}$ is a permutation matrix (with respect
to the standard basis $\ev{1},\ldots,\ev{r}$ of $\mathbb{C}^{r}$).
\end{description}
Note that the permutation associated to $S^{2}$ ({\it charge conjugation})
is an involution $\mu\mapsto\overline{\mu}$, since $S^{4}=1$. Also,
$T$ has finite order dividing $N$ because of $\Gamma\left(N\right)<\ker\rho$,
and so is of the form $T=\exp\left(2\pi i\FB\right)$ for some
diagonal matrix\begin{equation}
\FB=\left(\begin{array}{ccc}
\ks_{1}\\
 & \ddots\\
 &  & \ks_{r}\end{array}\right)\,\,,\label{eq:expdef}\end{equation}
where $0\leq\ks_{1},\ldots,\ks_{r}<1$ are rational numbers - the
{\it exponents} of $\rho$ - whose denominator divides the level $N$. Because
$S^{2}$ commutes with $T$, one also has $\ks_{\overline{\mu}}=\ks_{\mu}$.
This means, denoting by $\ccorb$ the orbits of charge conjugation,
that the exponent is the same for each element of an orbit $\xi\in\ccorb$:
we'll denote by $\ks_{\xi}$ this common value. We'll also use the
notation $\ev{\xi}=\sum_{\mu\in\xi}\ev{\mu}$ for an orbit $\xi\in\ccorb$.

It is known that the above admissibility conditions are always satisfied
by the modular representation associated to a Rational Conformal Field
Theory. Indeed, the primary fields of the RCFT provide a distinguished
basis in which $T$ is diagonal and $S^{2}$ is a permutation matrix,
while the first admissibility condition follows from the results of \cite{Ban}.
More generally, an analogue will hold for certain (as yet undetermined)
classes of Vertex Operator Algebras and Modular Tensor Categories.

For an admissible representation $\rho:\SL\rightarrow\gl{r}{\mathbb{C}}$,
we'll denote by $\FD$ the space of vector-valued complex functions
$\FA:\FC\rightarrow\mathbb{C}^{r}$ which are holomorphic in the upper
half-plane $\FC=\set{\tau}{\mathrm{Im}\,\tau>0}$, satisfy the transformation
rule\begin{equation}
\FA\left(\frac{a\tau+b}{c\tau+d}\right)=\mr{a}{b}{c}{d}\FA\left(\tau\right)\label{eq:modtrans}\end{equation}
for all $\sm{a}{b}{c}{d}\!\in\!\SL$, and have only finite order poles
at the cusps. To explain this last condition, note that Eqs.(\ref{eq:modtrans})
and (\ref{eq:expdef}) imply that $q^{-\FB}\FA\left(\tau\right)$
is invariant under $\tau\mapsto\tau+1$, and so may be expanded into
a power series in $q=\exp\left(2\pi i\tau\right)$:\begin{equation}
q^{-\FB}\FA=\sum_{n\in\mathbb{Z}}\FA\left[n\right]q^{n}\,\,,\label{eq:q-series}\end{equation}
where $\FA\left[n\right]\!\in\!\mathbb{C}^{r}$ for all $n$. We define
\begin{equation}
\FE\FA=\sum_{n<0}\FA\left[n\right]q^{n}\label{eq:Pdef}\end{equation}
 to denote the sum of the negative powers of $q$ in Eq.(\ref{eq:q-series}),
i.e. the singular (or principal) part of $\FA$. Then the requirement of having only
finite order poles at the cusps means that each component of $\FE\FA$
is a polynomial in $q^{-1}$ for $\FA\!\in\!\FD$, i.e. we get a linear
map $\FE:\FD\rightarrow V$, where $V=\oplus_{\mu=1}^{r}\maxid\ev{\mu}$,
%and $\maxid$ is the maximal ideal of the polynomial algebra $\mathbb{C}\left[q^{-1}\right]$, i.e.
for the set $\maxid$ of all polynomials in $q^{-1}$ with vanishing constant
term. It follows from the basic principles of Rational Conformal Field
Theory, that the characters $\chi_\mu(\tau)$ of the primary fields form a vector that
belongs to the space $\FD$, where $\rho$ is the modular representation
associated to the given model (see \cite{zhu} for a more general statement). 
\footnote{We note that a closely related theory of vector-valued modular forms had been developed in \cite{KM1,KM2}.}

Let's note that Eq.(\ref{eq:modtrans}) implies \[
\FA\left(\tau\right)=\mr{-1}{0}{0}{-1}\FA\left(\tau\right)\,\,,\]
in other words any $\FA\left(\tau\right)\in\FD$ is invariant under
charge conjugation $S^{2}$. This means that the image of $\FE$ lies in the
subspace $V_{+}$ of $V$ whose elements are left invariant by $S^{2}$. Actually,
it follows from the Riemann-Roch theorem for vector bundles  \cite{mukai} (or the arguments on p.15--16 of
\cite{Gun}), that $\mathrm{im}\,\FE=V_{+}$,
since the (compactified) quotient  of the upper half-plane $\FC$ by $\SL$ is a sphere
(see \cite{BG} for details).
A basis of $V_{+}$ is provided by the vectors $q^{-m}\ev{\xi}$,
 for any positive integer $m$ and $S^2$-orbit $\xi\in\ccorb$.

The kernel of the map $\FE$ is also easy to describe. Indeed, suppose that
$\FA\in\FD$ has vanishing singular part: then all of its vector components
are holomorphic on the upper half-plane, including the cusps (since $\SL$
can map any cusp to $i\infty$). Since
the kernel of $\rho$ is a congruence subgroup, the components of
$\FA$ are holomorphic functions on the compact Riemann surface $\overline{\FC
/\ker\rho}$ uniformized
by $\ker\rho$, hence they are all constant. But this constant vector
$\FA$ should also satisfy Eq.(\ref{eq:modtrans}), and thus it should
belong to $\FF$, the invariant subspace of $\rho$. This shows that $\ker\FE=\FF$,
and is then finite dimensional (over $\mathbb{C}$).
Thus, the induced map $\FD/\FF\rightarrow V_{+}$ is a bijection.

The outcome of the above considerations is that any element $v\in V_{+}$
determines a unique coset $\FA\in\FD/\FF$ such that $\FE\FA=v$.
We'll denote by $\SB{\xi}{m}$ the coset for which $\FE\SB{\xi}{m}=q^{-m}\ev{\xi}$.
Since the $q^{-m}\ev{\xi}$ form a basis of $V_{+}$, the cosets $\SB{\xi}{m}$
form a basis of the linear space $\FD/\FF$, which we'll call the
{\it canonical basis}. Our main concern will be to determine explicitly
the $\SB{\xi}{m}$ for a given representation $\rho$.

Strictly speaking, the canonical basis
vectors $\SB{\xi}{m}$ are not elements of $\FD$, although they come close
to it, since their Laurent expansions are completely determined up
to addition of a constant term from $\FF$. To simplify the ensuing
presentation, we'll make the assumption that $\FF=0$: this small
loss of generality is amply compensated by the gain in clarity and
brevity. The general case can be worked out without too much effort.
Indeed, $W_\rho=0$ will generically hold, since $W_\rho\ne 0$ can only
occur when some $\lambda_\mu=0$ in Eq.(\ref{eq:expdef}). A familiar example
when $W_\rho\ne 0$ is provided by the trivial one-dimensional $\rho$ -
e.g.\ the $c=24$ theories with only 1 primary field have identical character,
up to an additive constant (which counts the number of spin-1 fields) running
from 0 to 1128 \cite{Sch}.

\section{The Hauptmodul and the recursion relations}

According to the results of the previous section, the space $\FD$
is an infinite dimensional linear space over $\mathbb{C}$. In particular,
the elements $\SB{\xi}{m}$ of the canonical basis are linearly independent,
and thus one needs seemingly an infinite amount of data to describe
the structure of $\FD$. As it turns out, the situation is much better,
and this is related to the existence of the Hauptmodul $J$ (or absolute
invariant) for $\SL$.

Indeed, when $\rho$ is the trivial one-dimensional representation
of $\SL$, the space $\FD$ is known \cite{Apostol,koblitz} to be the polynomial algebra
$\mathbb{C}\left[J\right]$, where $J$ is the Hauptmodul of $\SL$,
i.e. the unique holomorphic function on $\FC$ that is invariant under
the action of $\SL$, i.e. satisfies the functional equation\begin{equation}
J\left(\frac{a\tau+b}{c\tau+d}\right)=J\left(\tau\right)\label{eq:Jinv}\end{equation}
for all $\sm{a}{b}{c}{d}\in\SL$, and whose Laurent series around
$q=0$ reads \begin{equation}
J\left(\tau\right)=q^{-1}+\sum_{n=1}^{\infty}c\left(n\right)q^{n}\,\,.\label{eq:JLaurent}\end{equation}
We note that the coefficients $c\left(n\right)$ of the Laurent expansion
are all positive integers (since $J$ is the character of the Moonshine
module), equal to the dimensions of specific representations of the
Monster, the largest sporadic simple group \cite{flm}.

The importance of the Hauptmodul $J$ for our considerations stems
from the fact that for an admissible representation $\rho$ and any
$\FA\in\FD$, the product $J\FA$ is still an element of $\FD$. In
other words, the linear space $\FD$ is a module over the polynomial
algebra $\mathbb{C}\left[J\right]$. The all-important result is that
this module is finitely generated by the $\SB{\xi}{1}$-s. To see
this, let's consider the product $J\SB{\xi}{m}$. Since the map $\FE$
is linear, the knowledge of the Laurent expansions of $J$ and $\SB{\xi}{m}$
determines the singular part of the product, and this is enough to
determine $J\SB{\xi}{m}$ uniquely as a combination of the canonical
basis vectors. In particular, from Eq.(\ref{eq:JLaurent}) and the
Laurent expansion \begin{equation}
q^{-\FB}\SB{\xi}{m}=q^{-m}\ev{\xi}+\sum_{n=0}^{\infty}\SB{\xi}{m}\left[n\right]q^{n}\label{eq:sblaurent}\end{equation}
one gets \begin{equation}
\FE\left(J\SB{\xi}{m}\right)=\left(q^{-\left(m+1\right)}+\sum_{n=1}^{m-1}c\left(n\right)q^{n-m}\right)\ev{\xi}+q^{-1}\SB{\xi}{m}\left[0\right]\,\,,\label{eq:Jsbsing}\end{equation}
from which one reads off the following recursion relation

\begin{equation}
J\SB{\xi}{m}=\SB{\xi}{m+1}+\sum_{n=1}^{m-1}c\left(n\right)\SB{\xi}{m-n}+\sum_{\orb\in\ccorb}\frac{1}{\left|\orb\right|}\left\langle \SB{\xi}{m}\left[0\right],\ev{\orb}\right\rangle \SB{\orb}{1}\,\,,\label{eq:recursion}\end{equation}
where $\left|\orb\right|$ denotes the length of the orbit $\orb\in\ccorb$
(which is either 1 or 2, because charge conjugation is an involution).
This may be rearranged as \begin{equation}
\SB{\xi}{m+1}=J\SB{\xi}{m}-\sum_{n=1}^{m-1}c\left(n\right)\SB{\xi}{m-n}-\sum_{\orb\in\ccorb}\xm{\orb}{\xi}{m}\SB{\orb}{1}\,\,,\label{eq:recursion2}\end{equation}
where we have introduced the notation \begin{equation}
\xm{\orb}{\xi}{m}=\frac{1}{\left|\orb\right|}\left\langle \SB{\xi}{m}\left[0\right],\ev{\orb}\right\rangle \,\,.\label{xmdef}\end{equation}

Clearly, Eq.(\ref{eq:recursion2}) and the knowledge of the $\SB{\orb}{1}$-s for all $\orb\in\ccorb$
allows to compute the canonical basis
elements $\SB{\xi}{2}$, and inductively all $\SB{\xi}{m}$-s for
$m>1$, as linear combinations of the $\SB{\orb}{1}$-s, with coefficients
which are polynomials in $J$, proving the claim that $\FD$ is a
finitely generated $\mathbb{C}\left[J\right]$-module. Note that the
form of the recursion relations Eq.(\ref{eq:recursion2}) does not
depend explicitly on the representation $\rho$, only implicitly,
through the values $\xm{\eta}{\xi}{m}$ (of course, the latter are
determined by $\rho$).

\section{Eisenstein series and the differential relations}

It is a well known result of the theory of modular forms that some
differential operators map modular forms to modular forms. In the
present context, appropriate linear differential operators may be
found that map the space $\FD$ to itself, providing us with differential
relations between the different elements of the canonical basis.

To begin with, let's recall the definition of the (normalized) Eisenstein
series and the discriminant form \cite{Apostol,diamond,koblitz}. For a positive integer $k$ let
$\sigma_{k}\left(n\right)$ denote the sum of the $k$-th powers of
the divisors of the integer $n$, i.e.\begin{equation}
\sigma_{k}\left(n\right)=\sum_{d|n}d^{k}\,\,.\label{eq:sigmak}\end{equation}
The Eisenstein series of weight $2k$ is \begin{equation}
E_{2k}(\tau)=1-\frac{2k}{B_{k}}\sum_{n=1}^{\infty}\sigma_{2k-1}\left(n\right)q^{n}\,\,,\label{eq:eisensteindef}\end{equation}
where $B_{k}$ denotes the $k$-th Bernoulli number, and $q=\exp\left(2\pi i\tau\right)$
as usual. For
$1\le k\le 5$ the coefficient  $-2k/B_k$ equals $-24,+240,-504,+480,-264$,
respectively.

%small values of $k$ one gets \begin{subequations}
%
%\begin{eqnarray}
%E_{2}(\tau) & = & 1-24\sum_{n=1}\sigma_{1}\left(n\right)q^{n}\label{eq:eisexpl2}\\
%E_{4}(\tau) & = & 1+240\sum_{n=1}\sigma_{3}\left(n\right)q^{n}\label{eq:eisexpl4}\\
%E_{6})\tau) & = & 1-504\sum_{n=1}\sigma_{5}\left(n\right)q^{n}\label{eq:eisexpl6}\\
%E_{8}(\tau) & = & 1+480\sum_{n=1}\sigma_{7}\left(n\right)q^{n}\label{eq:eisexpl8}\\
%E_{10}(\tau) & = & 1-264\sum_{n=1}\sigma_{9}\left(n\right)q^{n}\label{eq:eisexpl10}\end{eqnarray}
%\end{subequations}

The Eisenstein series are holomorphic in the upper half-plane and
at the cusps, and for $k>1$ they are modular forms of weight $2k$,
i.e. they satisfy the transformation rule \begin{equation}
E_{2k}\left(\frac{a\tau+b}{c\tau+d}\right)=\left(c\tau+d\right)^{2k}E_{2k}\left(\tau\right)\,\:\label{eq:eistrans}\end{equation}
for all $\sm{a}{b}{c}{d}\in\SL$. There are many relations among the
Eisenstein series, e.g. \begin{subequations}\begin{eqnarray}
E_{8} & = & E_{4}^{2}\label{eq:eisrels1}\\
E_{10} & = & E_{4}E_{6}\,\,.\label{eq:eisrels2}\end{eqnarray}
\end{subequations}

The discriminant form $\Delta$ is defined by the infinite product
\begin{equation}
\Delta(\tau)=q\prod_{n=1}^{\infty}\left(1-q^{n}\right)^{24}\,\,.\label{eq:discrdef}\end{equation}
It is a cusp form of weight 12, i.e. it is holomorphic in the upper
half-plane, has a (first order) zero at the cusp $\tau=i\infty$,
and satisfies \begin{equation}
\Delta\left(\frac{a\tau+b}{c\tau+d}\right)=\left(c\tau+d\right)^{12}\Delta\left(\tau\right)\label{eq:discrtrans}\end{equation}
for all $\sm{a}{b}{c}{d}\in\SL$.  It is related
to the Eisenstein series $E_{4}$ and $E_{6}$ by the formula \begin{equation}
1728\Delta=E_{4}^{3}-E_{6}^{2}\,\,.\label{eq:discrrel}\end{equation}
An important property of $\Delta$
is that it doesn't vanish on the upper half-plane $\FC$.

% While the Eisenstein series $E_{2}$ fails to be a modular form, it
% is nevertheless closely related to them, since it is proportional
% to the logarithmic derivative of the discriminant form $\Delta$:
% \begin{equation}
% \frac{d\Delta}{d\tau}=2\pi i\Delta E_{2}\,\,.\label{eq:discrE2}\end{equation}
Finally, let's note that the Hauptmodul $J$ may be expressed through
the above quantities as\begin{equation}
J=\frac{E_{4}^{3}}{\Delta}-744\,\,.\label{eq:Jdiscr}\end{equation}

Let's consider the linear differential operator

\begin{equation}
\DO=\frac{E_{10}}{2\pi i\Delta}\frac{d}{d\tau}\,\,.\label{eq:nabladef}\end{equation}
This operator maps the space $\FD$ to itself. Indeed, since the ratio
$E_{10}/\Delta$ is holomorphic in $\FC$, with a first order pole
at $\tau=i\infty$, it follows that $\DO\FA$ is also holomorphic
in $\FC$ and has only finite order poles at $\tau=i\infty$ for $\FA\in\FD$.
On the other hand, differentiating both sides of Eq.(\ref{eq:modtrans}),
and taking into account the transformation rules Eqs.(\ref{eq:eistrans})
and (\ref{eq:discrtrans}), one gets \[
\DO\FA\left(\frac{a\tau+b}{c\tau+d}\right)=\mr{a}{b}{c}{d}\DO\FA\left(\tau\right)\,\,,\]
showing that $\DO\FA\in\FD$, as claimed.

This result  allows to derive (first order) differential relations
between the canonical basis vectors $\SB{\xi}{m}$. To illustrate
this point, let's apply $\DO$ to the Laurent expansion Eq.(\ref{eq:sblaurent})
of the canonical basis vector $\SB{\xi}{m}$. A straightforward computation
shows that \begin{equation}
\FE\left(\DO\SB{\xi}{m}\right)=\left(\ks_{\xi}-m\right)\sum_{n=-1}^{m-1}\ei_{n}q^{n-m}\ev{\xi}+q^{-1}\sum_{\orb\in\ccorb}\ks_{\orb}\xm{\orb}{\xi}{m}\ev{\orb}\,\,,\label{eq:nablasing}\end{equation}
where $\xm{\orb}{\xi}{m}$ is the constant matrix introduced in Eq.(\ref{xmdef}),
and the integers $\ei_{n}$ are the Laurent coefficients of $E_{10}/\Delta$:
\begin{equation}
\frac{E_{10}}{\Delta}=\sum_{n=-1}^{\infty}\ei_{n}q^{n}\,\,=q^{-1}-240-141444q-8529280q^{2}+\ldots\,\,.\label{eq:eidef}\end{equation}
Since the singular part determines the whole series, it follows from
Eq.(\ref{eq:nablasing}) that \begin{equation}
\DO\SB{\xi}{m}=\left(\ks_{\xi}-m\right)\sum_{n=-1}^{m-1}\ei_{n}\SB{\xi}{m-n}+\sum_{\orb\in\ccorb}\ks_{\orb}\xm{\orb}{\xi}{m}\SB{\orb}{1}\,\,.\label{eq:difrel}\end{equation}

In particular, for $m=1$ Eq.(\ref{eq:difrel}) reduces to\begin{equation}
\DO\SB{\xi}{1}=\left(\ks_{\xi}-1\right)\left(\SB{\xi}{2}-240\SB{\xi}{1}\right)+\sum_{\orb\in\ccorb}\ks_{\orb}\xm{\orb}{\xi}{1}\SB{\orb}{1}\,\,.\label{eq:difrel1}\end{equation}
But $\SB{\xi}{2}$ is determined by the recursion relation Eq.(\ref{eq:recursion2}),
as a linear combination of the $\SB{\orb}{1}$-s. Substituting its
expression into Eq.(\ref{eq:difrel1}) leads to \begin{equation}
\DO\SB{\xi}{1}=\left(\ks_{\xi}-1\right)\left(J-240\right)\SB{\xi}{1}+\sum_{\orb\in\ccorb}\left(1+\ks_{\orb}-\ks_{\xi}\right)\xm{\orb}{\xi}{1}\SB{\orb}{1}\,\,.\label{eq:difeq}\end{equation}
Thus we get a system of first order linear differential equations
satisfied by the canonical basis vectors $\SB{\xi}{1}$, which may
be rewritten as \begin{equation}
\frac{1}{2\pi i}\frac{d\SB{\xi}{1}}{d\tau}=\sum_{\orb}\mathcal{D}_{\orb}^{\xi}\left(q\right)\SB{\orb}{1}\,\:\label{eq:difeq2}\end{equation}
upon introducing the square matrix \begin{equation}
\mathcal{D}_{\orb}^{\xi}\left(q\right)=\frac{\Delta}{E_{10}}\left\{ \left(J-240\right)\left(\ks_{\xi}-1\right)\delta_{\orb}^{\xi}+\left(1+\ks_{\orb}-\ks_{\xi}\right)\xm{\orb}{\xi}{1}\right\} \,\,.\label{eq:Dmatdef}\end{equation}
Note that this matrix is meromorphic in the upper half-plane (it has
first order poles at $\tau=\exp\left(2\pi i/3\right)$ and $\tau=i$),
and holomorphic at the cusp $\tau=i\infty$, i.e. it has a Laurent
expansion \begin{equation}
\mathcal{D}_{\orb}^{\xi}\left(q\right)=\sum_{n=0}^{\infty}\mathcal{D}_{\orb}^{\xi}\left[n\right]q^{n}\label{eq:Dmatexp}\end{equation}
without negative powers of $q$. The first few coefficients of the
above expansion read \begin{eqnarray}
\mathcal{D}_{\orb}^{\xi}\left[0\right] & = & \left(\ks_{\xi}-1\right)\delta_{\orb}^{\xi}\nonumber \\
\mathcal{D}_{\orb}^{\xi}\left[1\right] & = & \left(1+\ks_{\orb}-\ks_{\xi}\right)\xm{\orb}{\xi}{1}\label{eq:Dmatcoeffs}\\
\mathcal{D}_{\orb}^{\xi}\left[2\right] & = & 338328\left(\ks_{\xi}-1\right)\delta_{\orb}^{\xi}+240\left(1+\ks_{\orb}-\ks_{\xi}\right)\xm{\orb}{\xi}{1}\,\,.\nonumber \end{eqnarray}
Differential equations obeyed by the characters of Rational Conformal Field
Theories have been studied elsewhere, where they have been used for
e.g. classification purposes and studying modularity (see  \cite{mms,zhu}, resp.).

The differential equation Eq.(\ref{eq:difeq2}), supplemented with
the boundary conditions $\FE\left(\SB{\xi}{1}\right)=q^{-1}\ev{\xi}$
and $\SB{\xi}{1}\left[0\right]=\sum_{\orb}\xm{\orb}{\xi}{1}\ev{\orb}$,
allows in principle to determine recursively the coefficients $\SB{\xi}{1}\left[n\right]$
for $n>0$. Indeed, substituting into Eq.(\ref{eq:difeq2}) the expansions
Eqs.(\ref{eq:Dmatexp}) and (\ref{eq:sblaurent}), and taking into
account that $\frac{1}{2\pi i}\frac{d}{d\tau}=q\frac{d}{dq}$, one
gets, after comparing powers of $q$, the relation

\begin{equation}
\left(n+1-\ks_{\xi}+\FB\right)\SB{\xi}{1}\left[n\right]=\sum_{\orb\in\ccorb}\left(\mathcal{D}_{\orb}^{\xi}\left[n+1\right]\ev{\orb}+\sum_{m=1}^{n}\mathcal{D}_{\orb}^{\xi}\left[m\right]\SB{\orb}{1}\left[n-m\right]\right)\,\,.\label{eq:difrecur}\end{equation}
For example, after some rearrangements the above formula gives for
$n=1$\[
\SB{\xi}{1}\left[1\right]=\sum_{\orb}\frac{\mathcal{D}_{\orb}^{\xi}\left[2\right]+\sum_{\nu}\left(1+\ks_{\nu}-\ks_{\xi}\right)\xm{\nu}{\xi}{1}\xm{\orb}{\nu}{1}}{2+\ks_{\orb}-\ks_{\xi}}\ev{\orb}\,\,,\]
and similar expressions may be obtained for the higher terms. This
means that the knowledge of the exponents and of the matrix $\xm{\orb}{\xi}{1}$
of Eq.(\ref{xmdef}) allows the explicit computation of the canonical basis vectors $\SB{\xi}{1}$,
and hence - via the recursion relations Eq.(\ref{eq:recursion2})
- of all the elements of the canonical basis.

\section{Invariants and covariants}

The notion of invariants and covariants (aka. equivariant polynomial
maps) will play an important role in what follows, so let's sketch
their definition. Consider the polynomial algebra in $r$ variables
$R=\mathbb{C}\left[x_{1},\ldots,x_{r}\right]$: to each matrix $A\!\in\!\gl{r}{\mathbb{C}}$
is associated the algebra map \begin{eqnarray}
\hat{A}:R & \rightarrow & R\label{polact}\\
x_{i} & \mapsto & \sum_{j}A_{ij}x_{j}\,\,.\nonumber \end{eqnarray}

If $G$ is a subgroup of $\gl{r}{\mathbb{C}}$, an invariant of $G$
is a polynomial $\mathcal{I}\!\in\! R$ left fixed by $\hat{A}$ for
all $A\!\in\! G$. The invariants of $G$ form a subring\begin{equation}
\inv=\set{\mathcal{I}\in R}{\hat{A}\left(\mathcal{I}\right)=\mathcal{I}\,\,\mathrm{for}\,\,\mathrm{all}\,\, A\in G}\,\,\label{eq:invring}\end{equation}
of the algebra $R$, which inherits the natural grading from $R$.
It is known that under mild conditions (e.g. if $G$ is linearly reductive)
the algebra $\inv$ is finitely generated \cite{mukai}.

A covariant of the subgroup $G<\gl{r}{\mathbb{C}}$ is an algebra
map $\phi:R\rightarrow R$ such that $\hat{A}\circ\phi=\phi\circ\hat{A}$
for all $A\!\in\! G$. The set of covariants of $G$ is graded by
degree (as algebra maps) and, besides forming a linear space over $\mathbb{C}$,
is a (graded) $\inv$ module, since for a covariant $\phi$ and an
invariant $\mathcal{I}$ the map $\mathcal{I}\phi:x_{i}\mapsto\mathcal{I}\phi\left(x_{i}\right)$
is again a covariant of $G$.

Covariants of degree 0 are related to the commutant of $G$: indeed,
if $\phi$ is a covariant of degree 0, then there exists an $r$-by-$r$
matrix $M$ such that $\phi\left(x_{i}\right)=\sum_{j}M_{ij}x_{j}$
(because $\phi$ has degree 0), and $AM=MA$ for all $A\!\in\! G$
(because $\phi$ is a covariant). Conversely, for a matrix $M$ in
the commutant of $G$, the map $\phi:x_{i}\mapsto\sum_{j}M_{ij}x_{j}$
is a covariant of degree 0.

By an invariant (covariant) of a finite dimensional matrix representation
$\rho$ of $\SL$, we'll mean an invariant (resp. covariant) of the
image of $\rho$. The importance of covariants stems from the fact
that for $\FA\!\in\!\FD$ and $\phi$ a covariant of $\rho$, one
has $\phi\left(\FA\right)\!\in\!\FD$, since $\phi\left(\FA\right)$
is holomorphic - being a polynomial expression in holomorphic functions
-, has only finite order poles at the cusps, and transforms according
to the representation $\rho$. Similarly, if $\mathcal{I}$ is an
invariant and $\FA\!\in\!\FD$, then $\mathcal{I}\left(\FA\right)\!\in\!\mathbb{C}\left[J\right]$,
because it is holomorphic, has only finite order poles at the cusps,
and is invariant under all modular transformations.

\section{A worked-out example: the Yang-Lee model}

In the present section we'll illustrate, on the example of the Yang-Lee
model, how the results of the previous sections may be used to determine
explicitly the elements $\SB{\xi}{m}$ of the canonical basis of $\FD/\FF$,
and hence all solutions to Eq.(\ref{eq:modtrans}) holomorphic in $\FC$ and
meromorphic at the cusps.

The Yang-Lee model is the Virasoro minimal model $M\left(5,2\right)$
of central charge $c=-\frac{22}{5}$. Its exponents are \[
\left(\begin{array}{c}
\ks_{1}\\
\ks_{2}\end{array}\right)=\frac{1}{60}\left(\begin{array}{c}
11\\
59\end{array}\right)\,\,,\]
while its S matrix reads \[
S=\sqrt{\frac{2}{5+\sqrt{5}}}\left(\begin{array}{cc}
1 & \frac{1+\sqrt{5}}{2}\\
\frac{1+\sqrt{5}}{2} & -1\end{array}\right)\,\,.\]
The corresponding representation $\rho$ is admissible (since it comes
from an RCFT), has trivial charge conjugation, and trivial invariant subspace
$\FF=0$.

The map \begin{eqnarray}
\phi:\mathbb{C}\left[x_{1},x_{2}\right] & \rightarrow & \mathbb{C}\left[x_{1},x_{2}\right]\label{eq:lycov}\\
\left(\begin{array}{c}
x_{1}\\
x_{2}\end{array}\right) & \mapsto & \left(\begin{array}{c}
p\left(x_{1},x_{2}\right)\\
p\left(x_{2},-x_{1}\right)\end{array}\right)\nonumber \end{eqnarray}
is a covariant of degree 48, where \begin{multline}
p\left(x_{1},x_{2}\right)=x_{2}^{49}-114464x_{1}^{10}x_{2}^{39}-1586424x_{1}^{15}x_{2}^{34}-4273878x_{1}^{20}x_{2}^{29}\\
+3491397x_{1}^{25}x_{2}^{24}-559580x^{30}x_{2}^{19}+952812x_{1}^{35}x_{2}^{14}-14063x_{1}^{40}x_{2}^{9}+294x_{1}^{45}x_{2}^{4},\label{eq:covpoly}\end{multline}
while \begin{equation}
\mathcal{I}=x_{1}x_{2}\left(x_{1}^{10}+11x_{1}^{5}x_{2}^{5}-x_{2}^{10}\right)\label{eq:lyinv}\end{equation}
is an invariant of degree 12.

Let's turn to the computation of the canonical basis. By definition,
$\SB{2}{1}$ has a Laurent expansion of the form\begin{equation}
\SB{2}{1}(\tau)=q^{\FB}\left(\begin{array}{c}
a_{0}+a_{1}q+\ldots\\
q^{-1}+b_{0}+b_{1}q+\ldots\end{array}\right)\,\,,\label{eq:ly21exp}\end{equation}
where the coefficients $a_{0},b_{0},\ldots$ have to be determined.
Inserting this expression into Eq.(\ref{eq:lyinv}), we have \begin{equation}
\mathbb{\mathcal{I}}\left(\SB{2}{1}\right)=-a_{0}+\left(11a_{0}^{6}-11b_{0}a_{0}-a_{1}\right)q+\ldots\label{eq:i21exp}\end{equation}
This means that $\mathcal{I}\left(\SB{2}{1}\right)$ is holomorphic
at $q=0$, hence it should be a constant, i.e. the coefficients of
the positive powers of $q$ in the expansion Eq.(\ref{eq:i21exp})
should vanish%
\footnote{Note that $a_{0}=0$ is not possible, since Eq.(\ref{eq:lyinv}) would 
then imply that the first component of $\SB{2}{1}$
vanishes identically, which is incompatible with Eq.(\ref{eq:modtrans}).%
}. In particular, one has \begin{eqnarray}
a_{1} & = & 11a_{0}\left(a_{0}^{5}-b_{0}\right)\,\,,\label{eq:a1}\\
a_{2} & = & 727a_{0}^{11}-781b_{0}a_{0}^{6}+(66b_{0}^{2}-11b_{1})a_{0}\,\,.\label{eq:a2}\end{eqnarray}

On the other hand, because $\phi$ is a covariant, we know that $\phi\left(\SB{2}{1}\right)$
belongs to $\FD$. Inserting Eq.(\ref{eq:ly21exp}) into $\phi$,
one gets \begin{equation}
\phi\left(\SB{2}{1}\right)=q^{\FB}\left(\begin{array}{c}
q^{-1}+49b_{0}+(1176b_{0}^{2}+49b_{1}-114464a_{0}^{10})q+\ldots\\
294a_{0}^{4}q^{-1}+26999a_{0}^{9}+294b_{0}a_{0}^{4}+\ldots\end{array}\right)\,\,,\label{lyp21exp}\end{equation}
from which one reads off\begin{equation}
\phi\left(\SB{2}{1}\right)=\SB{1}{1}+294a_{0}^{4}\SB{2}{1}\,\,\label{lyp21}\end{equation}
by comparing the singular parts. This leads at once to the following
$q$-expansion, taking into account Eq.(\ref{eq:a1}) \begin{equation}
\SB{1}{1}(\tau)=q^{\FB}\left(\begin{array}{c}
q^{-1}+49b_{0}-294a_{0}^{5}+\left(1176b_{0}^{2}+49b_{1}-117698a_{0}^{10}+3234b_{0}a_{0}^{5}\right)q+\ldots\\
26999a_{0}^{9}+\left(3413445a_{0}^{14}-1592941b_{0}a_{0}^{9}\right)q+\ldots\end{array}\right)\,\,.\label{ly11exp}\end{equation}

Note that from the above we can read off the entries of the matrix of
Eq.(\ref{xmdef}):
\begin{equation}
\xm{\eta}{\xi}{1}=\left(\begin{array}{cc}
49(b_{0}-6a_{0}^{5}) & 26999a_{0}^{9}\\
a_{0} & b_{0}\end{array}\right)\,\,.\label{lyxmat}\end{equation}

This shows that, should we know the values of $a_{0}$ and $b_{0}$,
we could compute the $q$-expansion of the $\SB{\xi}{1}$-s recursively
via Eq.(\ref{eq:difrecur}). To determine these parameters, let's
plug the expression Eq.(\ref{eq:ly21exp}) into Eq.(\ref{eq:difeq}),
and equate the coefficients of the constant terms on each side (the
coefficients of the negative $q$ powers are equal by construction),
which gives

\begin{subequations}

\begin{eqnarray}
72a_{0}^{6}-(24b_{0}+48)a_{0} & = & 0\,\,,\label{eq:lyd21A}\\
-\frac{26999}{5}a_{0}^{10}-b_{0}^{2}-240b_{0}+2b_{1}+\frac{28194}{5} & = & 0\,\,,\label{eq:lyd21B}\end{eqnarray}
\end{subequations}which can be solved to give (recall that $a_{0}\neq0$)
\begin{subequations}\begin{eqnarray}
b_{0} & = & 3a_{0}^{5}-2\,\,,\label{eq:lyb0}\\
b_{1} & = & \frac{1}{5}\left(13522a_{0}^{10}+3540a_{0}^{5}-15287\right)\,\,.\label{eq:lyb1}\end{eqnarray}
\end{subequations}Equating the coefficients of the terms linear in
$q$, and taking into account Eqs.(\ref{eq:lyb0}) and (\ref{eq:lyb1}),
one gets \begin{eqnarray}
-\frac{373248}{5}a_{0}^{11}+\frac{373248}{5}a_{0} & = & 0\,\,,\label{eq:lyd21C}\\
\frac{1324812}{5}a_{0}^{15}-\frac{9654708}{5}a_{0}^{10}-\frac{2837277}{5}a_{0}^{5}+3b_{2}+\frac{11167158}{5} & = & 0\,\,,\label{eq:lyd21D}\end{eqnarray}
from which follows that $a_{0}^{10}=1$, i.e. $a_{0}$ is a tenth
root of unity. The other expansion coefficients may be expressed in
terms of $a_{0}$: \begin{eqnarray}
b_{0} & = & 3a_{0}^{5}-2\nonumber \\
a_{1} & = & 22a_{0}(1-a_{0}^{5})\nonumber \\
b_{1} & = & 354a_{0}^{5}-353\label{eq:lycoeffs}\\
a_{2} & = & a_{0}(3125-3124a_{0}^{5})\nonumber \\
b_{2} & = & 100831a_{0}^{5}-100830\nonumber \end{eqnarray}
and so on.

All in all, we got 10 different possibilities for $\SB{2}{1}$, according
to the precise value of $a_{0}$. Only one of these does solve our
original problem, i.e. only one of them transforms according to the
representation $\rho$: it can be selected by e.g. determining the
corresponding solution of the differential equation Eq.(\ref{eq:difeq2}),
and checking its transformation law under $\tau\mapsto-\frac{1}{\tau}$.
But in our case there is a shortcut: the character vector of the Yang-Lee
model has a first order pole in its second component, i.e. it equals
$\SB{2}{1}$, and being a character vector, its $q$-expansion coefficients
are all non-negative integers (being eigenvalue multiplicities). In
particular, $a_{0}$ should be a non-negative integer: the only 10-th
root of unity that satisfies this is $a_{0}=1$. Indeed, with this
value of $a_{0}$ we get the following $q$-expansion \begin{equation}
\SB{2}{1}(\tau)=q^{\FB}\left(\begin{array}{c}
1+q^{2}+\ldots\\
q^{-1}+1+q+q^{2}+\ldots\end{array}\right)\,\,,\label{eq:lychar}\end{equation}
recovering the well-known result for the character vector of the Yang-Lee
model (see e.g. \cite{rc}). From this and Eq.(\ref{ly11exp}) one gets \begin{equation}
\SB{1}{1}(\tau)=q^{\FB}\left(\begin{array}{c}
q^{-1}-245-113239q-6029989q^{2}+\ldots\\
26999+1820504q+\ldots\end{array}\right)\,\,.\label{eq:lyx11}\end{equation}
Note that this cannot be the character vector of a RCFT, since some
of its expansion coefficients are negative (though still integers). This
shows that the modular representation in an RCFT does constrain the singular
part $\FE\FA$ of character vectors (e.g. the conformal weights $h_\mu$)
in a nontrivial way. These constraints go far beyond the inequality
$\sum_\mu(h_\mu-c/24)\le r(r-1)/12$ of \cite{mms}, which is satisfied by
Eq.(\ref{eq:lyx11}) and indeed by any $\FA\in\FD$.

This arbitrariness up to a 10-th root of unity is not surprising in hindsight,
and certainly does not contradict our earlier claim that $\ker\FE=W_\rho$ (which vanishes
here). The only ingredients which went into Eqs.(\ref{eq:lycoeffs}) and the
constraint $a_0^{10}=1$ were the exponents $\lambda_\mu$, the covariant
$\phi$ in Eq.(\ref{eq:lycov}), and invariant $\mathcal{I}$ in Eq.(\ref{eq:lyinv}).
Of course these are all determined by the modular representation $\rho$, but
it is easy to find other modular representations having the same $\lambda_\mu,
\phi,\mathcal{I}$ which correspond to the 9 other values of $a_0$.

\section{A second example: the Ising model}

The Ising model is the Virasoro minimal model $M(4,3)$ of central charge $c=\frac{1}{2}$.
It has 3 primary fields, and the modular representation $\rho$ associated
to it is characterized by \[
S=\frac{1}{2}\left(\begin{array}{ccc}
1 & 1 & \sqrt{2}\\
1 & 1 & -\sqrt{2}\\
\sqrt{2} & -\sqrt{2} & 0\end{array}\right)\]
and the exponents \[
\left(\begin{array}{c}
\lambda_{1}\\
\lambda_{2}\\
\lambda_{3}\end{array}\right)=\frac{1}{48}\left(\begin{array}{c}
47\\
23\\
2\end{array}\right)\,\,.\]
Note that $\rho$ - besides being admissible - is irreducible, has
trivial charge conjugation, and $\FF=0$.

The map \begin{eqnarray}
\phi_{n}:\mathbb{C}\left[x_{1},x_{2},x_{3}\right] & \rightarrow & \mathbb{C}\left[x_{1},x_{2},x_{3}\right]\nonumber \\
\left(\begin{array}{c}
x_{1}\\
x_{2}\\
x_{3}\end{array}\right) & \mapsto & \frac{1}{2}\left(\begin{array}{c}
\left(x_{1}+x_{2}\right)^{24n+1}+\left(-1\right)^{n}\left(x_{1}-x_{2}\right)^{24n+1}\\
\left(x_{1}+x_{2}\right)^{24n+1}-\left(-1\right)^{n}\left(x_{1}-x_{2}\right)^{24n+1}\\
\left(-1\right)^{n}2^{12n+1}x_{3}^{24n+1}\end{array}\right)\label{eq:isingcov}\end{eqnarray}
is a covariant (of degree $24n$) for any non-negative integer $n$.

The above information is already enough to determine the canonical
basis of $\FD$ along the lines presented in the previous section.
Instead of going through this lengthy calculation, which doesn't present
any difficulties, we'll exploit the fact that the character vector
of the Ising model is known (see e.g. \cite{rc}):\begin{equation}
\FA_{\textrm{Ising}}=\frac{1}{2}\left(\begin{array}{c}
\w{}+\w{1}\\
\w{}-\w{1}\\
\sqrt{2}\w{2}\end{array}\right)\,\,,\label{eq:isingchar}\end{equation}
where \begin{eqnarray}
\mathfrak{f}(\tau) & = & q^{-1/48}\prod_{n=0}^{\infty}\left(1+q^{n+\frac{1}{2}}\right)\,\,,\nonumber \\
\mathfrak{f}_{1}(\tau) & = & q^{-1/48}\prod_{n=0}^{\infty}\left(1-q^{n+\frac{1}{2}}\right)\,\,,\label{eq:weberdef}\\
\mathfrak{f}_{2}(\tau) & = & \sqrt{2}q^{1/24}\prod_{n=1}^{\infty}\left(1+q^{n}\right)\,\,\nonumber \end{eqnarray}
are the Weber functions. Note that, while linearly independent, the
Weber functions are not algebraically independent, for they satisfy
the identities

\begin{subequations}\begin{align}
\mathfrak{f}_{1}^{8}+\mathfrak{f}_{2}^{8} & =\mathfrak{f}^{8}\,\,,\label{eq:weberrel1}\\
\w{}\w{1}\w{2} & =\sqrt{2}\,\,.\label{eq:weberrel2}\end{align}
\end{subequations}Moreover, they are related to the Hauptmodul through\begin{equation}
J+744=\frac{\left(\mathfrak{f}^{24}-16\right)^{3}}{\mathfrak{f}^{24}}=\frac{\left(\mathfrak{f}_{1}^{24}+16\right)^{3}}{\mathfrak{f}_{1}^{24}}=\frac{\left(\mathfrak{f}_{2}^{24}+16\right)^{3}}{\mathfrak{f}_{2}^{24}}\,\,.\label{eq:weberJ}\end{equation}

$\FA_{\textrm{Ising}}$ has a pole of order 1 in its first component,
which means that $\FA_{\textrm{Ising}}=\SB{1}{1}$, i.e. \begin{equation}
\SB{1}{1}=\frac{1}{2}\left(\begin{array}{c}
\w{}+\w{1}\\
\w{}-\w{1}\\
\sqrt{2}\w{2}\end{array}\right)\,\,.\label{eq:isx11}\end{equation}

Applying $\phi_{1}$ and $\phi_{2}$ to $\FA_{\textrm{Ising}}$, and
comparing the singular parts, one gets the relations \begin{eqnarray}
\phi_{1}\left(\FA_{\mathrm{Ising}}\right) & = & 25\SB{1}{1}+\SB{2}{1}\,\,,\label{eq:isphi1}\\
\phi_{2}\left(\FA_{\mathrm{Ising}}\right) & = & \SB{1}{2}+1176\SB{1}{1}+49\SB{2}{1}\,\,,\label{eq:isphi2}\end{eqnarray}
from which one deduces \begin{equation}
\SB{2}{1}=\frac{1}{2}\left(\begin{array}{c}
\w{}^{25}-\w{1}^{25}-25\w{}-25\w{1}\\
\\\w{}^{25}+\w{1}^{25}-25\w{}+25\w{1}\\
\\-\sqrt{2}\w{2}\left(25+\w{2}^{24}\right)\end{array}\right)\label{eq:isx21}\end{equation}
and \begin{equation}
\SB{1}{2}=\frac{1}{2}\left(\begin{array}{c}
\w{}^{49}-49\w{}^{25}+49\w{}+\w{1}^{49}+49\w{1}^{25}+49\w{1}\\
\\\w{}^{49}-49\w{}^{25}+49\w{}-\w{1}^{49}-49\w{1}^{25}-49\w{1}\\
\\\sqrt{2}\w{2}\left(49+49\w{2}^{24}+\w{2}^{48}\right)\end{array}\right)\,\,.\label{eq:isx12}\end{equation}

Finally, from the recursion relation \begin{equation}
J\SB{1}{1}=\SB{1}{2}+\SB{2}{1}+\SB{3}{1}\,\,,\label{eq:isrecur}\end{equation}
 taking into account the relations Eqs.(\ref{eq:weberrel1}), (\ref{eq:weberrel2})
and (\ref{eq:weberJ}), one computes\begin{equation}
\SB{3}{1}=\left(\begin{array}{c}
8\w{}^{17}\w{1}^{8}-8\w{}^{24}\w{1}-128\w{}+\frac{\w{2}^{7}}{\sqrt{2}}\left(\w{}^{39}-\w{1}^{39}-16\w{}^{15}-32\w{1}^{15}\right)\\
\\8\w{}^{17}\w{1}^{8}+8\w{}^{24}\w{1}-128\w{}-\frac{\w{2}^{7}}{\sqrt{2}}\left(\w{}^{39}+\w{1}^{39}-16\w{}^{15}+32\w{1}^{15}\right)\\
\\\w{}^{15}\w{1}^{7}\left(\w{}^{24}-16\right)-8\sqrt{2}\w{2}\w{}^{24}\end{array}\right)\,\,.\label{eq:isx31}\end{equation}
 Thus, we have been able to determine explicitly the elements $\SB{\xi}{1}$
of the canonical basis. Note that from these one may derive explicit
expressions for the $\SB{\xi}{m}$ with $m>1$ by using the recursion
relation Eq.(\ref{eq:recursion2}).

Finally, from the above explicit expressions one gets the $q$-expansions

\begin{equation}
\SB{1}{1}(\tau)=q^{\FB}\left(\begin{array}{c}
q^{-1}+q+q^{2}+2q^{3}+\ldots\\
1+q+q^{2}+q^{3}+\ldots\\
1+q+q^{2}+2q^{3}+\ldots\end{array}\right)\,\,,\label{eq:isx11exp}\end{equation}

\begin{equation}
\SB{2}{1}(\tau)=q^{\FB}\left(\begin{array}{c}
2325+60630q+811950q^{2}+7502125q^{3}+\ldots\\
q^{-1}+275+13250q+235500q^{2}+2558550q^{3}+\\
-25-4121q-102425q^{2}-1331250q^{3}+\ldots\end{array}\ldots\right)\,\,,\label{eq:isx21exp}\end{equation}
and

\begin{equation}
\SB{3}{1}(\tau)=q^{\FB}\left(\begin{array}{c}
94208+9515008q+356765696q^{2}+7853461504q^{3}+\ldots\\
-4096-1130496q-63401984q^{2}-1763102720q^{3}+\ldots\\
q^{-1}-23+253q-1794q^{2}+9384q^{3}+\ldots\end{array}\right)\,\,.\label{eq:isx31exp}\end{equation}
Of these, only $\SB{1}{1}$ may be the character vector of a RCFT,
since it is the only one whose $q$-expansion coefficients are all
non-negative integers, illustrating again that the singular parts
of character vectors are heavily constrained.

\section{Further questions and developments}

This paper explains to what extent the $\SL$ representation $\rho$ determines
the vector-valued complex function $\FA$, and how in practise to construct it. 
This study suggests a number of additional questions.

We illustrate with examples how to construct the canonical basis vectors
$\FA^{(\eta;1)}$ coefficient by coefficient, and in principle the differential
equation Eq.(\ref{eq:difeq2}) tells us the full series.
% Well, in my opinion you should reconsider the next two sentences
But is it possible to express these $\FA^{(\eta;1)}$ using
known transcendental functions, much as we did with the Ising model (and more
generally has been done with all the minimal and Wess-Zumino-Witten models)? A reason
to think we can is that each component of $\FA^{(\eta;1)}$ will be a modular function
for some $\Gamma(N)$, and all of these can be expressed in terms of the $J$
function and $N^2$ Fricke functions $f_{r,s}$ (see e.g.\ \cite{La}).
An alternate approach was followed in \cite{ES}, who in five specific Conformal
Field Theoretic models explained how to write the characters using theta
functions, by relating $\rho$ to Weil representations; their method should
be quite general.

We are most interested in the modular representation $\rho$ and character
vector $\FA$ coming from Rational Conformal Field Theory. In this case
the components $\chi_\mu(\tau)$ of $\FA$ have a $q$-expansion whose coefficients
are non-negative integers. This is quite special; what are the
properties of $\rho$ which makes this possible? Given such a $\rho$,
which vectors in $V_+$ will be the principal parts of such non-negative
integer $\FA$? Integrality is easy to understand, using Galois methods.
In particular, for any admissible $\rho$, some positive integer multiple of each
canonical basis vector $\FA^{(\eta;m)}$ (translating by a vector in $W_\rho$ if
necessary) will have integer $q$-expansions, provided
all entries of the matrices $\rho(A)$, for all $A\in\SL$, lie in the cyclotomic
field ${\mathbb Q}[e^{2\pi i/N}]$, and in addition the $\ell$-th Galois
automorphism, for all $\ell$ coprime to $N$, applied entry by entry to the
matrix $S$, equals $\rho\sm{\ell}{0}{0}{\ell^{-1}}S$
% \left(\begin{matrix}{\ell&0\cr 0&\ell^{-1}}\end{matrix}\right)\,S$
(see \cite{BG} for more details and the proof).
 These conditions are automatically satisfied in any Rational Conformal
Field Theory \cite{Ban}.
Positivity of those $q$-expansions seems more difficult to understand, although
it is easy to verify that, unless $S$ has a strictly positive eigenvector
with eigenvalue 1,  no $\FA\in\FD$ can have a non-negative $q$-expansion.
As the examples in sections 6 and 7 illustrate, non-negativity is subtle and
would be very interesting to understand.

Curiously, in all examples we've seen, the $q$-expansions of the
canonical basis vectors $\FA^{(\eta;1)}$ have been either completely
non-negative, completely nonpositive, or alternating in sign (apart from
the $q^{-1}$ and constant terms, in some cases). Is this a general
phenomenon?

The analysis of the covariants and invariants for holomorphic orbifolds,
or equivalently the modular representations coming from quantum-doubles of
finite groups, is straightforward, and thus they supply a large family of
examples which can be worked out quite explicitly \cite{BG}.

Some modular representations $\rho$ (involving certain powers of 2) are more
exceptional than others \cite{NW}. Are these in any way special from the
point of view considered here? Can those exceptional modular representations
be realized in a Rational Conformal Field Theory?

\section*{Acknowledgments}
P.B. would like to thank Geoffrey Mason for stimulating discussions.
The work of P.B. was supported by research grants OTKA T047041, T037674, 
T043582, TS044839, the J\'anos Bolyai Research Scholarship of the Hungarian 
Academy of Sciences, and EC Marie Curie RTN, MRTN-CT-2004-512194. 
T.G. would like to thank E\"otv\"os University
and the University of Hamburg for kind hospitality while this research was
undertaken. His research is supported in part by NSERC and the Humboldt
Foundation.

\end{document}